\documentclass[review]{elsarticle}
\biboptions{numbers,sort&compress}
\usepackage{lineno,hyperref}
\usepackage{amsmath}
\usepackage{graphicx}
\usepackage{caption}
\usepackage{graphicx}
\usepackage{epstopdf}
\usepackage{caption}
\usepackage{subfigure}
\usepackage{float}
\modulolinenumbers[5]
\DeclareGraphicsExtensions{.eps,.ps,.jpg,.bmp}
\journal{Journal of \LaTeX\ Templates}

\begin{document}

\begin{frontmatter}

\title{Entangling cavity modes in a double-cavity optomechanical system}

\author{Ding-Shan Liu}
\author{Pu-Tong Wang}%
\author{Ming Jin}%
\author{Miao Yin$^*$}%
\address{School of Physics and Optoelectronics, South China University of Technology, Guangzhou 510640, People's Republic of China}

\cortext[mycorrespondingauthor]{Corresponding author. Fax: +86 20 87113934.}
\ead{scmyin@scut.edu.cn (M. Yin)}

\begin{abstract}
We study entanglement of the cavity modes in a double-cavity optomechanical system in strong-coupling regime. The system consists of two optomechanical systems coupled by a single photon hopping between them. With the radiation pressure of the photon, entanglement of the cavity modes can be generated. The concurrence between the cavity modes is at least twice larger than that between the mechanical modes. Moreover, when we change the ratio between coupling strength and resonant frequency of mechanical modes, the entanglement in cavity and mechanical modes are influenced differently.

\end{abstract}


\end{frontmatter}

\linenumbers

\section{Introduction}
Quantum entanglement\cite{entan}, as an important bridge between classical and quantum mechanics, has received extensive concern in different branches of physics.
With its intriguing features in microscopic world\cite{jiuchanzongshu}, many applications can be realized, such as quantum teleportation\cite{yinxingct}, quantum computing\cite{liangzijisuan}, quantum information\cite{inforbook}, and even the measurement of gravitational wave\cite{yinlibo}.
What is more, it is also important to study quantum entanglement in macroscopic systems. A lot of schemes are proposed, including entanglement between mirrors\cite{guangyajiuchan,sanjiaojing}, atomic ions\cite{naturejiuchanzhendang}, and cooper pairs\cite{weiqiangkubo}. It is also demonstrated that macroscopic entanglement can be generated via radiation pressure in optomechanical systems.

Cavity optomechancis\cite{qiangguangzongshu}, which focuses on the interation between light and mechanical oscillators, has become a more and more common approach to study entanglement in macroscopic systems with various schemes proposed\cite{qgjjiuchanzongshu,jingchangjiuchan,danjingdanqiang,prljixieyuqiangchang,01,djjg,qiangmojiuchan,yuanqiang,duoweijiuchan,zhenziliangzihua,1-28,E1,E2}.
For example, entanglement can be created between a cavity field and its output field\cite{jingchangjiuchan}, cavity field and a vibrating mirror inside\cite{prljixieyuqiangchang}, two distant cavities coupled with an optical fiber\cite{yuanqiang}, and two mechanical oscillators in coupled cavities\cite{E1}.

However, apart from some proposals of new models,
there has been progress in related technologies, including microfabrication\cite{nano}, cooling of mechanial motion \cite{laserjitailengque,2012lengque,prllengque}, experimental realization of strong coupling regime\cite{qiangouheshiyan} and single-photon regime\cite{danguangziqgj,sinphostr}.
Base on these, it is significant for us to conduct a further research on the entanglement in the single-photon strong-coupling regime.
Therefore, in this paper we investigate the entanglement of two cavity modes in a double-cavity optomechanical system. The system works in such a regime. It is found that entanglement can be generated apparently between the cavity modes. The concurrence between the cavity modes is at least twice larger than that between the mechanical modes. Additionally, we consider parameters appeared in the system and find different influences are made on entanglement between cavity modes and mechanical modes.

The outline of the paper is as follows. First, we describe a model of double-cavity optomechanic system and introduce the time-dependent state of the system. Second, we investigate the entanglement of cavity modes generated by state measurement.
Third, we discuss the differences and releationship between the entanglement of cavity modes and that of mechanical modes.  Finally, we summarize the main results of this work.

\section{Microcavity resonators and time-dependent state of the system}

As we known, many quantum effects in mechanical systems can be generated through radiation pressure. Especially in the single-photon strong-coupling regime, the system is sensitive enough that those effects can be produced by only one photon. In 2014, to generate quantum entanglement in this regime, J.Q.Liao et al.\cite{01} established a model of two-cavity optomechanical system, in which considerable entanglement between two mechanical modes is induced with a single photon hopping and providing	 the radiation pressure.

\begin{figure}
\centering
\includegraphics[width=10cm]{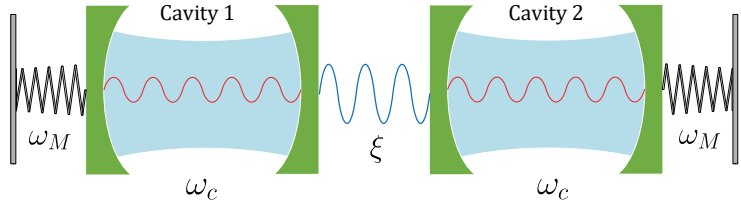}
\caption{Sketch of the studied system. The system is consist of two Fabry-Perot cavities, in each of the cavities there is a fixed mirror in the middle and a moving mirror on outer side. A single photon hopping between two cavities and provide radiation pressure to induce the entanglement.}
\label{fig1}
\end{figure}

In this paper, following the work of J.Q.Liao et al.\cite{01}, we consider entanglement between two cavity modes in the system.
As shown in Fig.\ref{fig1}, both two Fabry-Perot cavities in this system  is consist of a moving end mirror on one side and a fixed mirror on the other side. The single-mode fields in two cavities couple to the moving mirror by radiation pressure, and couple to each other through the hopping of a single photon. The system is working in the strong-coupling regime, so that the radiation pressure of the photon is considerable for interacting with the cavity.

In the analysis, the system is regarded as ideal, so that two cavities are identical and the photon decay of the cavity is neglected. Then as in Ref.\cite{01}, the Hamiltonian of the system is ($\hbar$= 1)

\begin{equation}
   \begin{aligned}
H_{S}=&\sum_{j=1,2}^{}\left[ {\omega_{c}c_{j}^{\dagger}c_{j}^{}+\omega_{M}m_{j}^{\dagger}m_{j}^{}-g_{0}c_{j}^{\dagger}c_{j}^{}(m_{j}^{\dagger}m_{j}^{})} \right]\\
      &-\xi (c_{1}^{\dagger}c_{2}^{}+c_{2}^{\dagger}c_{1}^{})
  \end{aligned}
\label{e2}
\end{equation}

Where $c_{j},c_{j}^{\dagger}$ and $m_{j},m_{j}^{\dagger}$ are the annihilation, creation operator of the single-mode cavity field and the mechanic mode of moving mirror in $j$th cavity, with their resonant frequencies of $\omega_{c}$ and $\omega_{M}$. $g_{0}$ is the single-photon coupling strength with $g_{0}=\frac{\omega_{c}x_{zpf}}{L}$, and $x_{zpf}=\sqrt{\frac{1}{2M\omega_{m}}}$ is the zero-point fluctuation of the moving mirror with mass M, and L is the rest length of the cavity. The parameter $\xi$ is the photon-hopping coupling strength between the two cavities.

On this basis, we introduce the unitary evolution operator corresponding to $H_{S}$, which is $U(t)=e^{-iH_{S}t}$, to figure out the state of the system at arbitrary time $t$. Since the system is working at single-photon regime, we set the initial state of the system is $|\psi(0)=|1\rangle_{c_{1}}|0\rangle_{a_{2}}|0\rangle_{c_{1}}|0\rangle_{m_{2}}$. After a series of complex computation, the expression of the time-dependent state under the rotating wave approximation can be derived as
\begin{equation}
   \begin{aligned}
|\Phi(t) \rangle=&U(t)|\Phi(0) \rangle
\\
=&\frac{1}{2}e^{i\Theta(t)}D_{m_{1}}[\beta(t)/\sqrt{2}]D_{m_{2}}[\beta(t)/\sqrt{2}]
\\
\times&\bigg( |1 \rangle_{c_{1}}|0 \rangle_{c_{2}} \bigg\{ [e^{i\omega_{M}t}+cos(gt)]|0 \rangle_{m_{1}} |0\rangle_{m_{2}}
\\
+&\frac{i}{\sqrt{2}}sin(gt)(|0 \rangle_{m_{1}}|1 \rangle_{m_{2}}-|1 \rangle_{m_{1}}|0 \rangle_{m_{2}}) \bigg\}
\\
+&|0 \rangle_{c_{1}}|1\rangle_{c_{2}} \bigg\{ [e^{i\omega_{M}t}-cos(gt)]|0 \rangle_{m_{1}} |0 \rangle_{m_{2}}
\\
+&\frac{i}{\sqrt{2}}sin(gt)(|0 \rangle_{m_{1}}|1 \rangle_{m_{2}}-|1 \rangle_{m_{1}}|0 \rangle_{m_{2}})  \bigg\} \bigg)
  \end{aligned}
\label{e3}
\end{equation}
Where $g=-\dfrac{g_{0}}{\sqrt{2}}$,
$\Theta(t)=\left(\omega_{c}+\frac{\omega_{M}}{2}\right)t+\dfrac{g^{2}}{\omega_{M}}\left[\dfrac{sin(\omega_{M}t)}{\omega_{M}}-t\right]$is the phace factor and $D_{m_{j}}[\beta(t)/\sqrt{2}]=
exp\left[ \frac{\beta(t)m_{j}^{\dagger}-\beta^{*}(t)m_{j}}{\sqrt{2}} \right]$
is the deplacement operator with $\beta(t)=-\dfrac{g}{\omega_{M}}\left(1-e^{-i\omega_{M}t}\right)$.

\section{Entanglement between cavity modes in the system}
Here, the whole system is in an entangled state, which involves cavity field mode $c_{1}$,$c_{2}$ and mechanical mode $m_{1}$,$m_{2}$. Therfore, we can produce entanglement between cavity field by measuring state of two mechanical modes. On the basis of equation \ref{e3}, here we list the possible results of the measuring:
\begin{enumerate}[(1)]
\item The mechanical mode was detected in the state of
$|0 \rangle_{m_{1}}|0 \rangle_{m_{2}}$, which possibility is $P_{00}=\frac{1+cos^2(gt)}{2}$.
Then the cavity field must be
\begin{equation}
   \begin{aligned}
|\phi_{00}(t) \rangle=D_{m_{1}}[\beta(t)/\sqrt{2}]D_{m_{2}}[\beta(t)/\sqrt{2}]|\psi_{00}(t) \rangle
   \end{aligned}
\label{e4}
\end{equation}
where
\begin{equation}
   \begin{aligned}
|\psi_{00}(t)\rangle=A_{00}\bigg\{|1\rangle_{c_{1}}|0 \rangle_{c_{2}}\big[e^{i\omega_{M}t}+cos(gt)\big]+|0\rangle_{c_{1}}|1 \rangle_{c_{2}}\big[e^{i\omega_{M}t}-cos(gt)\big]\bigg\}
   \end{aligned}
\label{e5}
\end{equation}
With the normalization factor $A_{00}=\frac{1}{\sqrt{2+2cos^{2}(gt)}}$.

Here, in order to measure the degree of the entanglement, we introduce the concurrence proposed by Wootters\cite{jcpj} as the critierion, which is
\begin{equation}
C(\psi)=
\left|{\langle\psi|\widetilde{\psi} \rangle}\right|
\label{e7}
\end{equation}
where
\begin{equation}
|\widetilde{\psi}\rangle=\sigma_{y}|\widetilde{\psi}^{*}\rangle, \  \sigma_{y}=
\bigg(
  \begin{array}{cc}
    0 & -i\\
    i & 0\\
  \end{array}
\bigg)
\label{e8}
\end{equation}
We plug $|\phi_{00}(t) \rangle$ in the function, then
\begin{equation}
   \begin{aligned}
C_{00}=&\left|{\langle\psi_{00}|\sigma_{y}|\widetilde{\psi_{00}}^{*}\rangle}\right|
\\
=& \left| \frac{e^{-2i\omega t}-cos^{2}(gt)}{1+cos^{2}(gt)} \right|
\\
=& \frac{\sqrt{1-2cos^2(gt)cos(2\omega t)+cos^4(gt)}}{1+cos^2(gt)}
   \end{aligned}
\label{e9}
\end{equation}

\item The mechanical mode was detected in the state of
$|0 \rangle_{m_{1}}|1 \rangle_{m_{2}}$ or $|1 \rangle_{m_{1}}|0 \rangle_{m_{2}}$, which possibilitys are  $P_{01}=P_{10}=\frac{sin^2(gt)}{4}$. Then the cavity field must be:

\begin{subequations}
 \begin{align}
|\phi_{01}(t) \rangle=D_{m_{1}}[\beta(t)/\sqrt{2}]D_{m_{2}}[\beta(t)/\sqrt{2}]|\psi_{01}(t) \rangle
\\
|\phi_{10}(t) \rangle=D_{m_{1}}[\beta(t)/\sqrt{2}]D_{m_{2}}[\beta(t)/\sqrt{2}]|\psi_{10}(t) \rangle
 \end{align}
\end{subequations}

and
\begin{subequations}
 \begin{align}
&|\phi_{01}(t) \rangle=A_{01}\left\{\frac{i}{\sqrt{2}}sin(gt)\big(\left|0 \right\rangle_{m_{1}}\left|1 \right\rangle_{m_{2}}+\left|1 \right\rangle_{m_{1}}\left|0 \right\rangle_{m_{2}}\big)\right\}
\\
&|\phi_{10}(t) \rangle=A_{10}\left\{-\frac{i}{\sqrt{2}}sin(gt)\big(\left|0 \right\rangle_{m_{1}}\left|1 \right\rangle_{m_{2}}+\left|1 \right\rangle_{m_{1}}\left|0 \right\rangle_{m_{2}}\big)\right\}
\label{e10}
 \end{align}
\end{subequations}
With the normalization factor $A_{01}=A_{10}=\frac{1}{sin(gt)}$.
And their concurrences is
\begin{equation}
C_{01}=C_{10}=1
\label{e11}
\end{equation}
\end{enumerate}

\section{Discussions}

\begin{figure}[H]
\includegraphics[width=1.0\textwidth]{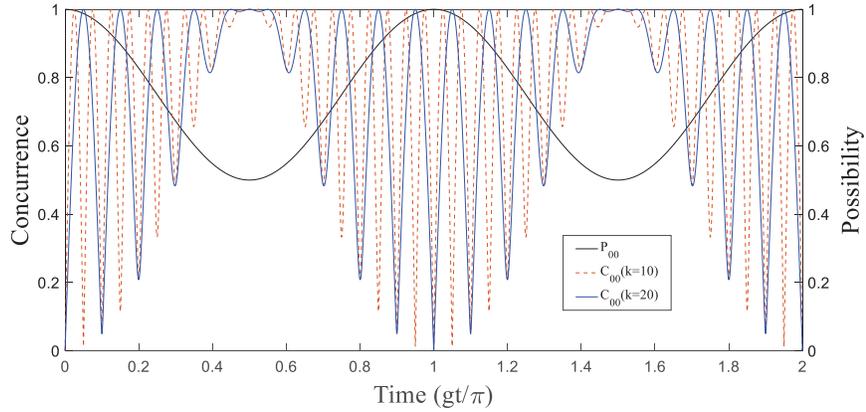}\caption{Evolution of the concurrence $C_{00}$ considering two values of $k=\frac{w}{g}$, including $k=10$ (blue-soild line) and $k=20$ (red-dash line), as well as their
 possibilty $P_{1}$(black-soild line). These values bring oscillations with different frequencies in their envelope, but all of them share the same possibilty $P_{00}$.}
\label{fig3}
\end{figure}

In Fig.\ref{fig3}, the concurrence $C_{00}$ of two cavity modes in case of $|0\rangle_{m1}|0\rangle_{m2}$ is demonstrated with scaled time $gt$. In our derivation above, we considered the case $\omega_{M}\gg g$ for rotating wave approximation. In this figure, we consider two values of $k=\frac{\omega}{g}$, including $k=10$ and $k=20$. We can find the concurrence oscillate periodically with sinusoidal envelopes, with their frequencies in the envelop various from $k$.
That is different from the concurrences in mechanical modes. It was given from previous research on mechanical modes\cite{01} that the concurrence and its corresponding possibility is (take $C_{1}$ and $P_{1}$ as example, in which the cavity mode is detected in $|1 \rangle_{c_{1}}|0 \rangle_{c_{2}}$)

\begin{subequations}
 \begin{align}
&C_{1}(t)=\frac{sin^2(gt)}{2[1+cos(gt)cos(\omega_{M}t)]}
\\
&P_{1}(t)=\frac{1+cos(gt)cos(\omega_{M}t)}{2}
\label{e12}
 \end{align}
\end{subequations}

\begin{figure}[H]
\centering
{\includegraphics[width = 0.9\linewidth]{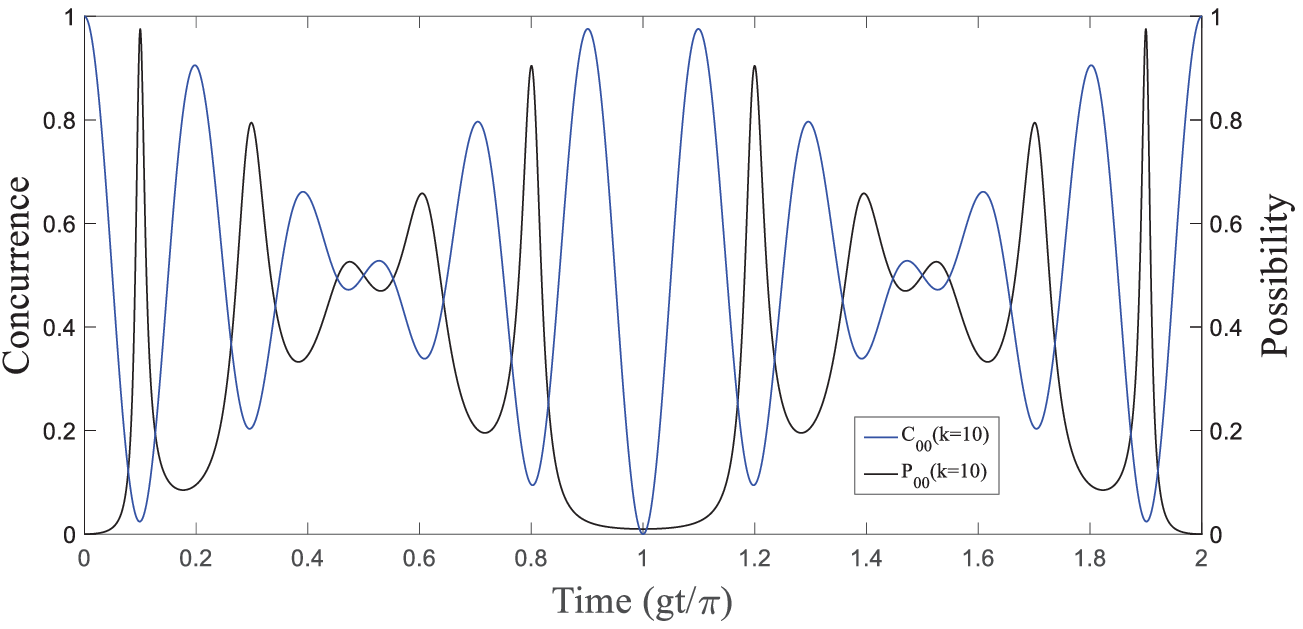}}\
{\includegraphics[width = 0.9\linewidth]{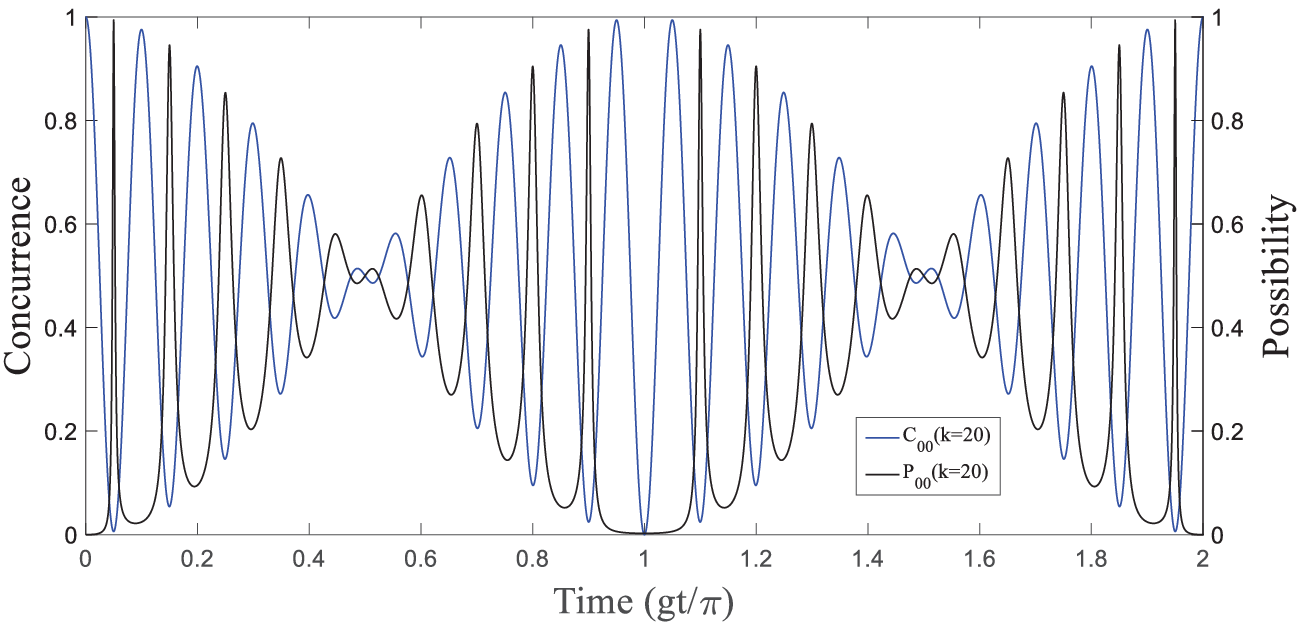}}\
\caption{Evolution of the concurrence $C_{1}$ in mechanical modes, considering two values of $k=\frac{w}{g}$, including $k=10$ (blue-soild line, upper figure) and $k=20$ (blue-soild line, lower figure). Unlike concurrence in cavity modes, their possibilties $P_{1}$ (black-soild line) various from different values of $k$.}
\label{e13}
\end{figure}

As shown in Fig.\ref{e13}, oscillation of the concurrence keep pace with its corresponding possibilty. When the concurrence oscillate more frequently, its possibility act the same, keeping negative correlation between them.

Since the concurrences and their corresponding possibilities are both oscillating, the average concurrence is induced for futher discussion, to describe the ability to generate entanglement of the system better:
\begin{equation}
C_{ave}=C(t)P(t).
\label{e14}
\end{equation}

For mechanic modes, there is
\begin{subequations}
\begin{align}
C_{ave}(mechanical)=&P_{1}C_{1}+P_{2}C_{2}=\frac{sin^2(gt)}{2}
\\
C_{ave}(cavity)=\,&P_{00}C_{00}+P_{01}C_{01}+P_{10}C_{10}
\\
=&\frac{\sqrt{1-2cos^2(gt)cos(2\omega t)+cos^4(gt)}}{1+cos^2(gt)}+\frac{sin^2(gt)}{2}
\end{align}
\label{e15}
\end{subequations}

Specially, for cavity modes in state of $|\psi_{00}\rangle$, there is
\begin{equation}
\begin{aligned}
C_{ave}(cavity,|\psi_{00}\rangle)=\,&P_{00}C_{00}
\\
=&\frac{\sqrt{1-2cos^2(gt)cos(2\omega t)+cos^4(gt)}}{2}
\end{aligned}
\label{e17}
\end{equation}

\begin{figure}[H]
\includegraphics[width=1.2\textwidth]{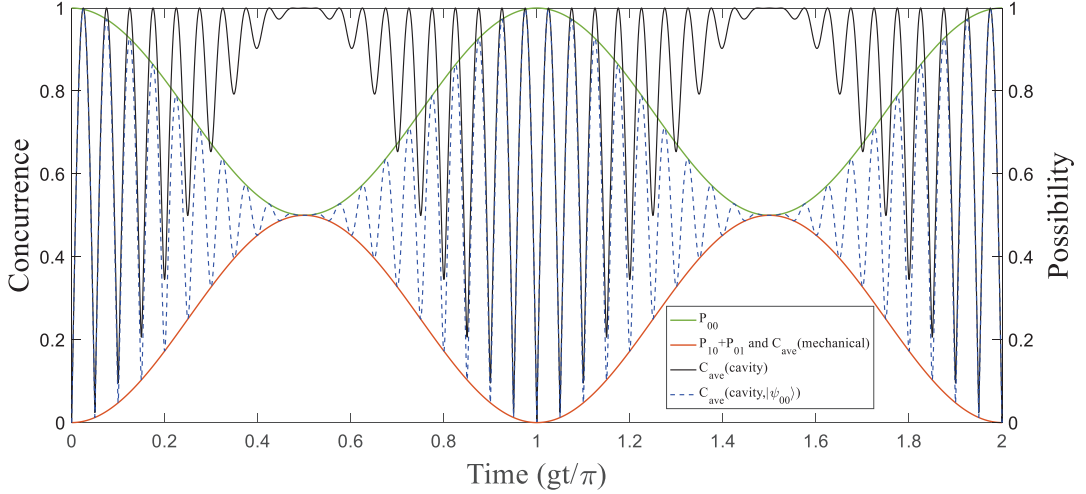}\caption{Comparison between concurrences in cavity modes and mechanical modes. $P_{00}$ is the possibility of the state $|\psi_{00}\rangle$ (green-soild line). $C_{ave} (mechanical)$ is the average concurrence of mechancal modes, which is also equal to $P_{10}+P_{01}$ (brown-soild line).
They consist the envelop of $C_{ave} (cavity,|\psi_{00}\rangle)$, which is the average concurrence of cavity modes in the state $|\psi_{00}\rangle$(blue-dashed line). Above them is the average concurrence of cavitiy modes $C_{ave} (cavity)$, represented by black-soild line.}
\label{fig4}
\end{figure}

Fig \ref{fig4} clearly shows the dynamics of quantities that related to our discussion on average concurrence. 
From formula.\ref{e3}, we can see there are three possibile states of mechanical modes ($|0 \rangle_{m_{1}}|0 \rangle_{m_{2}}$,$|1 \rangle_{m_{1}}|0 \rangle_{m_{2}}$ and $|0 \rangle_{m_{1}}|1 \rangle_{m_{2}}$), whose corresponding possibilties are $P_{00}, P_{10}$, $P_{01}$ and they add up to $1$.
Besides the fast oscillation of $C_{ave}$, which is correspond to photon's motion, we can find there is a slow oscillation of $P_{00}$ and $P_{10}+P_{01}$, which is correspond to the energy exchenge between mechanical modes and the light field.

When $P_{00}$ is large, the photon is oscillating in two cavities, with the mechanical system remained in the state $|0 \rangle_{m_{1}}|0 \rangle_{m_{2}}$ and not been entangled. There is a superposition state of $|0 \rangle_{c_{1}}|1 \rangle_{c_{2}}$ and $|1 \rangle_{c_{1}}|0 \rangle_{c_{2}}$. In this case, the concurrence of cavity modes can change from $0$ to $1$ freely, and no concurrence in mechanical modes.

When $P_{00}$ is small, the photon is entangled with the mechanical system while the mechanical modes can support the superposition. 
Because of the entanglement with mechanical system, the motion of the photon is not as free as that when $P_{00}$ is large, and in this case the detection of $|0 \rangle_{m_{1}}|0 \rangle_{m_{2}}$ provides more information about the photon. 
This make $C_{00} \rightarrow1$ and its average concurrence $C_{ave} (cavity,|\psi_{00}\rangle)\rightarrow P_{00}$. 
Therefore, we have the condition of greatest concurrence: $C_{ave} (cavity)\rightarrow1$ and $C_{ave} (mechanical)\rightarrow\frac{1}{2}$.
 
Meanwhile, when we consider the concurrence of mechanical modes after detection of cavity modes, only $|1 \rangle_{m_{1}}|0 \rangle_{m_{2}}$ and $|0 \rangle_{m_{1}}|1 \rangle_{m_{2}}$ contribute to the concurrence with the same coefficient. Hence the expectation of concurrence $C_{ave} (mechanical)=P_{01}+P_{10}$. 
This is also equal to $C_{ave} (cavity,|\psi_{01}\rangle)+C_{ave} (cavity,|\psi_{10}\rangle)$ because $C_{01}=C_{10}=1$. 
Besides, since $C_{ave} (cavity,|\psi_{00}\rangle)$ is oscillating in the envelop of $P_{00}$ and $P_{10}+P_{01}$, we always have $C_{ave} (cavity,|\psi_{00}\rangle)\geq C_{ave}(mechanical)$.

So as the sum of the average concurrence of $|0 \rangle_{m_{1}}|0 \rangle_{m_{2}}$, $|1 \rangle_{m_{1}}|0 \rangle_{m_{2}}$ and $|0 \rangle_{m_{1}}|1 \rangle_{m_{2}}$ , it is obvious that $C_{ave}(cavity)\geq 2 C_{ave}(mechanical)$. That is, we can generate the entanglement in cavity modes with its concurrence always at least twice of that in mechanical modes.

\section{Conclusions}
To sum up, we have investigated the entanglement of cavity modes in a double-cavity optomechanical system, based on the previous work of J.Q.Liao. It was found that with the radiation pressure of a single photon, the entanglement of cavity modes can be generated at least twice that of mechanical modes. Besides, the ratio $k=\frac{\omega}{g}$ makes different effects on cavity modes and mechanical modes. For cavity modes, the concurrence of $|\psi_{00}\rangle$ state and the average concurrence various from $k$ ,but the possibilty of detection act the same. For mechanical modes, the concurrence of its states and the corresponding possibitly various from $k$, but the average concurrence acts the same.

\section*{Acknowledgments}
We would like to express our gratitude to Prof.Dan M.Stamper-Kurn for his valuable discussion and advice to our work. 

This work is supported by the National Natural Science Foundation of China(No.11204088), Science and Technology Program of Guangzhou, China(No.201607010019), Science and Technology Planning Project of Guangdong Province, China(No.2015B010127004), and SRP of South China University of Technology.

\bibliography{new}

\begin{thebibliography}{10}
\expandafter\ifx\csname url\endcsname\relax
  \def\url#1{\texttt{#1}}\fi
\expandafter\ifx\csname urlprefix\endcsname\relax\def\urlprefix{URL }\fi
\expandafter\ifx\csname href\endcsname\relax
  \def\href#1#2{#2} \def\path#1{#1}\fi

\bibitem{entan}
E.Schrödinger, Discussion of probability relations between separated systems,
  Mathematical Proceedings of the Cambridge Philosophical Society 31~(4) (1935)
  555--563.

\bibitem{jiuchanzongshu}
R.~Horodecki, P.~Horodecki, M.~Horodecki, K.~Horodecki, Quantum entanglement,
  Reviews of Modern Physics 81~(2) (2009) 865--942.

\bibitem{yinxingct}
C.~H. Bennett, G.~Brassard, C.~Crepeau, R.~Jozsa, A.~Peres, W.~K. Wootters,
  Teleporting an unknown quantum state via dual classical and
  einstein-podolsky-rosen channels, Phys Rev Lett 70~(13) (1993) 1895--1899.

\bibitem{liangzijisuan}
T.~Pellizzari, S.~A. Gardiner, J.~I. Cirac, P.~Zoller, Decoherence, continuous
  observation, and quantum computing: A cavity qed model, Phys Rev Lett 75~(21)
  (1995) 3788--3791.

\bibitem{inforbook}
J.~A. Jones, D.~D. Jaksch, Quantum Information, Computation and Communication,
  Cambridge University Press, 2012.

\bibitem{yinlibo}
A.~Buonanno, Y.~Chen, Signal recycled laser-interferometer gravitational-wave
  detectors as optical springs, Physical Review D 65~(4).

\bibitem{guangyajiuchan}
S.~Mancini, V.~Giovannetti, D.~Vitali, P.~Tombesi, Entangling macroscopic
  oscillators exploiting radiation pressure, Phys Rev Lett 88~(12) (2002)
  120401.

\bibitem{sanjiaojing}
M.~Pinard, A.~Dantan, D.~Vitali, O.~Arcizet, T.~Briant, A.~Heidmann, Entangling
  movable mirrors in a double-cavity system, Europhysics Letters (EPL) 72~(5)
  (2005) 747--753.

\bibitem{naturejiuchanzhendang}
J.~D. Jost, J.~P. Home, J.~M. Amini, D.~Hanneke, R.~Ozeri, C.~Langer, J.~J.
  Bollinger, D.~Leibfried, D.~J. Wineland, Entangled mechanical oscillators,
  Nature 459~(7247) (2009) 683--5.

\bibitem{weiqiangkubo}
A.~D. Armour, M.~P. Blencowe, K.~C. Schwab, Entanglement and decoherence of a
  micromechanical resonator via coupling to a cooper-pair box, Phys Rev Lett
  88~(14) (2002) 148301.

\bibitem{qiangguangzongshu}
M.~Aspelmeyer, T.~J. Kippenberg, F.~Marquardt, Cavity optomechanics, Reviews of
  Modern Physics 86~(4) (2014) 1391--1452.

\bibitem{qgjjiuchanzongshu}
M.~Brune, E.~Hagley, X.~Maître, G.~Nogues, C.~Wunderlich, J.~M. Raimond,
  S.~Haroche, Manipulating entanglement with atoms and photons in a cavity, in:
  Quantum Electronics Conference, 1998. IQEC 98. Technical Digest. Summaries of
  papers presented at the International, 2001, p. 145.

\bibitem{jingchangjiuchan}
C.~Genes, A.~Mari, P.~Tombesi, D.~Vitali, Robust entanglement of a
  micromechanical resonator with output optical fields, Physical Review A
  78~(3).

\bibitem{danjingdanqiang}
X.~Mi, J.~Bai, S.~Ke-hui, Robust entanglement between a movable mirror and a
  cavity field system with an optical parametric amplifier, The European
  Physical Journal D 67~(6).

\bibitem{prljixieyuqiangchang}
D.~Vitali, S.~Gigan, A.~Ferreira, H.~R. Bohm, P.~Tombesi, A.~Guerreiro,
  V.~Vedral, A.~Zeilinger, M.~Aspelmeyer, Optomechanical entanglement between a
  movable mirror and a cavity field, Phys Rev Lett 98~(3) (2007) 030405.

\bibitem{01}
J.-Q. Liao, Q.-Q. Wu, F.~Nori, Entangling two macroscopic mechanical mirrors in
  a two-cavity optomechanical system, Physical Review A 89~(1).

\bibitem{djjg}
W.~Ge, M.~Al-Amri, H.~Nha, M.~S. Zubairy, Entanglement of movable mirrors in a
  correlated-emission laser, Physical Review A 88~(2).

\bibitem{qiangmojiuchan}
M.~Ikram, F.~Saif, Engineering entanglement between two cavity modes, Physical
  Review A 66~(1).

\bibitem{yuanqiang}
C.~Joshi, J.~Larson, M.~Jonson, E.~Andersson, P.~Öhberg, Entanglement of
  distant optomechanical systems, Physical Review A 85~(3).

\bibitem{duoweijiuchan}
X.~B. Zou, K.~Pahlke, W.~Mathis, Creating the multidimensional entangled
  coherent states of two cavity modes, The European Physical Journal D 33~(2)
  (2005) 297--300.

\bibitem{zhenziliangzihua}
J.~Zhang, K.~Peng, S.~L. Braunstein, Quantum-state transfer from light to
  macroscopic oscillators, Physical Review A 68~(1).

\bibitem{1-28}
M.~Paternostro, D.~Vitali, S.~Gigan, M.~S. Kim, C.~Brukner, J.~Eisert,
  M.~Aspelmeyer, Creating and probing multipartite macroscopic entanglement
  with light, Phys Rev Lett 99~(25) (2007) 250401.

\bibitem{E1}
T.~Yousif, W.~Zhou, L.~Zhou, State transfer and entanglement of two mechanical
  oscillators in coupled cavity optomechanical system, Journal of Modern Optics
  61~(14) (2014) 1180--1186.

\bibitem{E2}
T.~Yousif, W.~Zhou, L.~Zhou, Bistability and entanglement of a two-mode cavity
  optomechanical system, International Journal of Theoretical Physics 55~(2)
  (2015) 901--910.

\bibitem{nano}
T.~J. Kippenberg, K.~J. Vahala, Cavity optomechanics: back-action at the
  mesoscale, Science 321~(5893) (2008) 1172--6.

\bibitem{laserjitailengque}
J.~Chan, T.~P. Alegre, A.~H. Safavi-Naeini, J.~T. Hill, A.~Krause,
  S.~Groblacher, M.~Aspelmeyer, O.~Painter, Laser cooling of a nanomechanical
  oscillator into its quantum ground state, Nature 478~(7367) (2011) 89--92.

\bibitem{2012lengque}
E.~Verhagen, S.~Deleglise, S.~Weis, A.~Schliesser, T.~J. Kippenberg,
  Quantum-coherent coupling of a mechanical oscillator to an optical cavity
  mode, Nature 482~(7383) (2012) 63--7.

\bibitem{prllengque}
I.~Wilson-Rae, N.~Nooshi, W.~Zwerger, T.~J. Kippenberg, Theory of ground state
  cooling of a mechanical oscillator using dynamical backaction, Phys Rev Lett
  99~(9) (2007) 093901.

\bibitem{qiangouheshiyan}
S.~Groblacher, K.~Hammerer, M.~R. Vanner, M.~Aspelmeyer, Observation of strong
  coupling between a micromechanical resonator and an optical cavity field,
  Nature 460~(7256) (2009) 724--7.

\bibitem{danguangziqgj}
A.~Nunnenkamp, K.~Borkje, S.~M. Girvin, Single-photon optomechanics, Phys Rev
  Lett 107~(6) (2011) 063602.

\bibitem{sinphostr}
U.~Akram, N.~Kiesel, M.~Aspelmeyer, G.~J. Milburn, Single-photon opto-mechanics
  in the strong coupling regime, New Journal of Physics 12~(8) (2010) 083030.

\bibitem{jcpj}
W.~K. Wootters, Entanglement of formation of an arbitrary state of two qubits,
  Physical Review Letters 80~(10) (1998) 2245--2248.

\end{thebibliography}

\end{document}